\documentstyle[psfig]{cupconf}

\ifoldfss
\else
  \ifnfssone
    \newmathalphabet{\mathit}
      \addtoversion{normal}{\mathit}{cmr}{m}{it}
      \addtoversion{bold}{\mathit}{cmr}{bx}{it}
    \newmathalphabet{\mathcal}
      \addtoversion{normal}{\mathcal}{cmsy}{m}{n}
    \else
    \ifnfsstwo
    \fi
  \fi
\fi

\def\hexnumber#1{\ifcase#1 0\or1\or2\or3\or4\or5\or6\or7\or8\or9\or
 A\or B\or C\or D\or E\or F\fi }

\makeatletter
\ifx\CUP@mtlplain@loaded\undefined
\else
\fi
\makeatother
\makeatletter
\ifx\CUP@mtlplain@loaded\undefined
\else
\fi
\makeatother
\title[H$_2$ in Star Forming Regions]{ISO Spectroscopy of H$_2$ in Star Forming Regions}

\author[M.E. van den Ancker {\it et al.}]
{M. E. van den Ancker$^{1,2}$,  P. R. Wesselius$^{3}$\\
\and \ns A. G. G. M. Tielens$^{3,4}$}

\affiliation{
$^1$Harvard-Smithsonian Center for Astrophysics, 
 Cambridge, MA 02138, USA\\[\affilskip]
$^2$University of Amsterdam, 
 Kruislaan 403, 1098 SJ  Amsterdam, The Netherlands\\[\affilskip]
$^3$SRON, P.O. Box 800, 9700 AV  Groningen, The Netherlands\\[\affilskip]
$^4$Kapteyn Astronomical Institute, P.O. Box 800, 
 9700 AV  Groningen, The Netherlands}

\begin{document}
\ifnfssone
\else
  \ifnfsstwo
  \else
    \ifoldfss
      \let\mathcal\cal
      \let\mathrm\rm
      \let\mathsf\sf
    \fi
  \fi
\fi

\maketitle

\begin{abstract}
We have studied molecular hydrogen emission in a sample of 21 YSOs using spectra 
obtained with the {\it Infrared Space Observatory} (ISO). 
H$_2$ emission was detected in 12 sources and can be 
explained as arising in either a shock, caused by the 
interaction of an outflow from an embedded YSO with the 
surrounding molecular cloud, or in a PDR surrounding 
an exposed young early-type star. The distinction between 
these two mechanisms can not always be made from the pure 
rotational H$_2$ lines alone. Other tracers, such as PAH 
emission or [S\,{\sc i}] 25.25~$\mu$m emission, are needed 
to identify the H$_2$ heating mechanism. No deviations from 
a 3:1 ortho/para ratio of H$_2$ were found. Both shocks 
and PDRs show a warm and a hot component in H$_2$, 
which we explain as thermal emission from warm molecular 
gas (warm component), or UV-pumped infrared fluorescence 
in the case of PDRs and the re-formation of H$_2$ for 
shocks (hot component).
\end{abstract}

\firstsection 
\section{Introduction}
Molecular hydrogen is expected to be ubiquitous in the circumstellar 
environment of Young Stellar Objects (YSOs). It is the main 
constituent of the molecular cloud from which the young star has 
formed and is also expected to be the main component of the 
circumstellar disk. Most of this material will be at temperatures 
of 20--30~K and difficult to observe. However, some regions 
may be heated to temperatures of a few hundred K and produce 
observable H$_2$ emission. The intense UV radiation generated by 
accretion as well as by the central star itself will create a 
photodissociation region (PDR), of which the surface layer
is heated by collisions with photoelectrically ejected electrons 
from grain surfaces, in any surrounding neutral material. 
Another possibility to produce warm H$_2$  
is in shocks caused by the interaction of an outflow with the surrounding 
molecular cloud. Shocks are usually divided into 
J- or Jump-shocks, and C- or Continuous-shocks. In J-shocks 
the molecular material is dissociated in the shock front, where the 
gas is heated to several times 10$^4$ degrees. Behind the shock front, 
molecular material will re-form, and warm H$_2$ may be observed 
in the post-shock gas. C-shocks, in contrast, are not sufficiently 
powerful to dissociate molecular material, but may produce observable 
amounts of H$_2$ within the shock front itself.
\begin{figure}
\centerline{\psfig{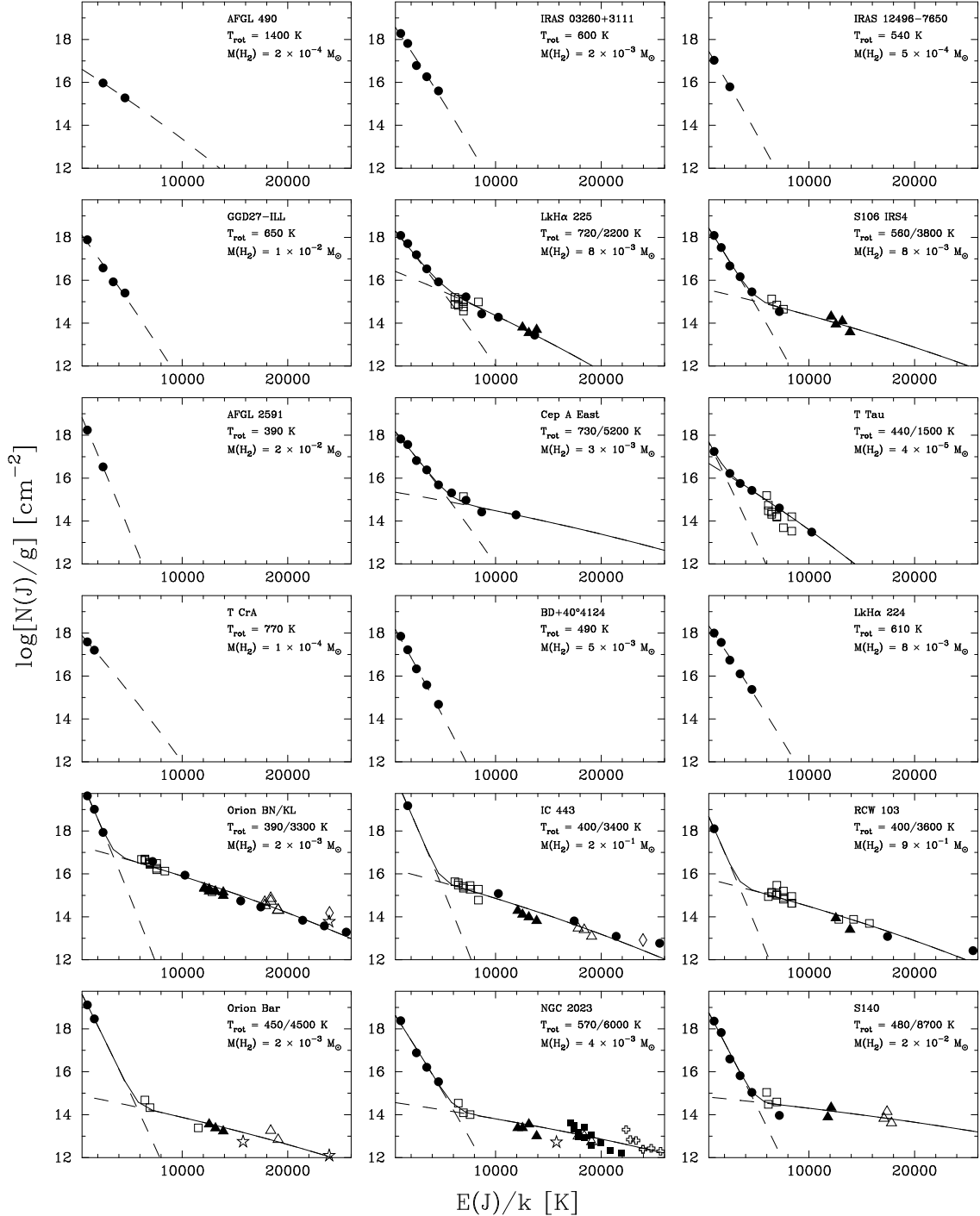}}
\caption[]{H$_2$ excitation diagrams for programme stars (top four rows) 
and comparison shocks and PDRs (bottom two rows). 
Shown are apparent columns of H$_2$ in the pure-rotational (0--0; filled 
dots), 1--0 (open squares), 2--1 (filled triangles), 2--0 (open stars), 
3--2 (open triangles), 3--0 (filled squares), 4--3 (open diamonds), 
and 4--1 transitions (open crosses). Observational errors are smaller 
than the size of the plot symbol. The Boltzmann distribution fits 
are plotted as dashed lines. The solid lines show the sum of both 
thermal components for each source.}
\end{figure}
\begin{figure}
\vspace*{0.1cm}
\centerline{\psfig{figure=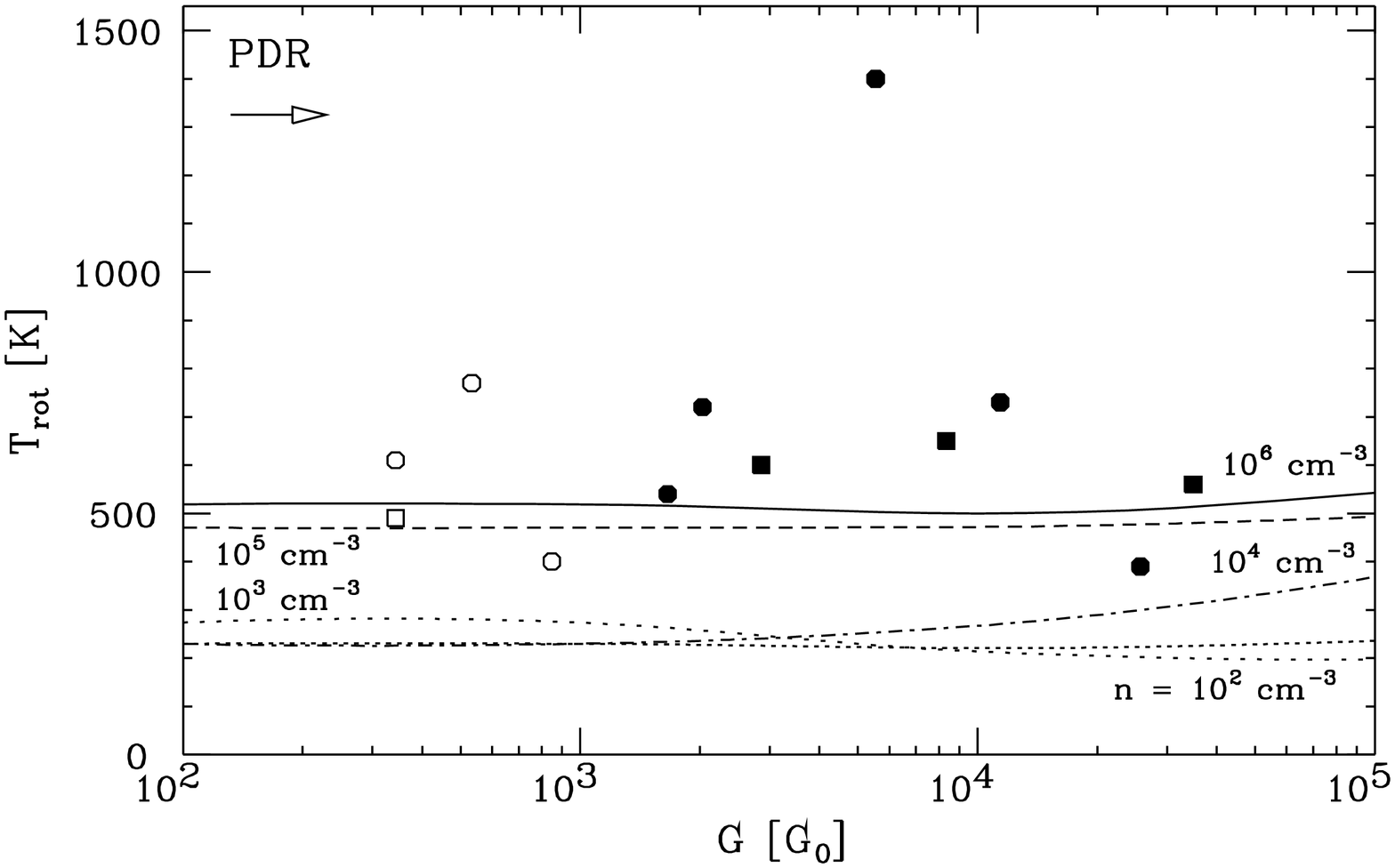,width=6.7cm,angle=90}
\hspace*{0.05cm}
\psfig{figure=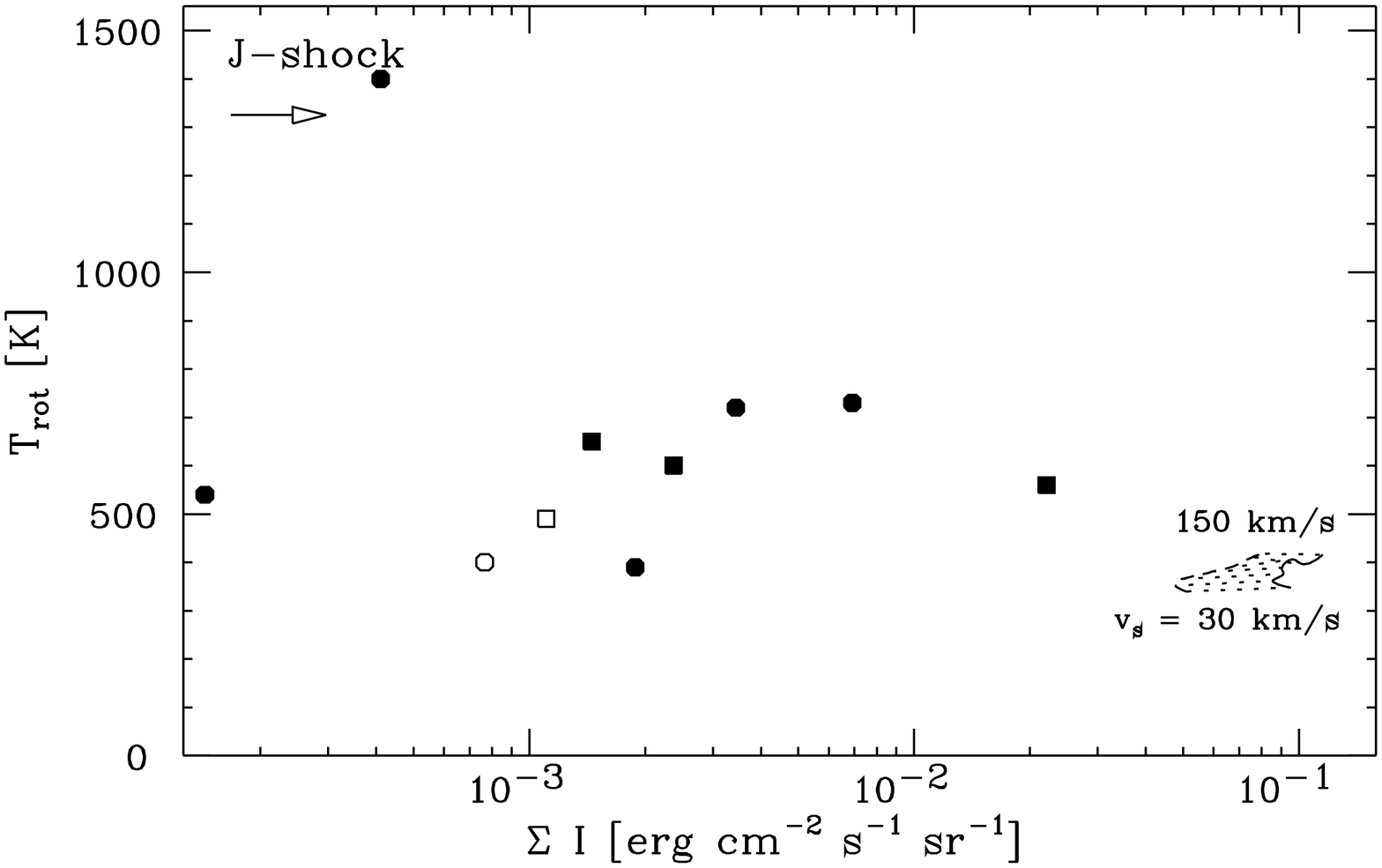,width=6.7cm,angle=90}}
\vspace*{0.1cm}
\centerline{\psfig{figure=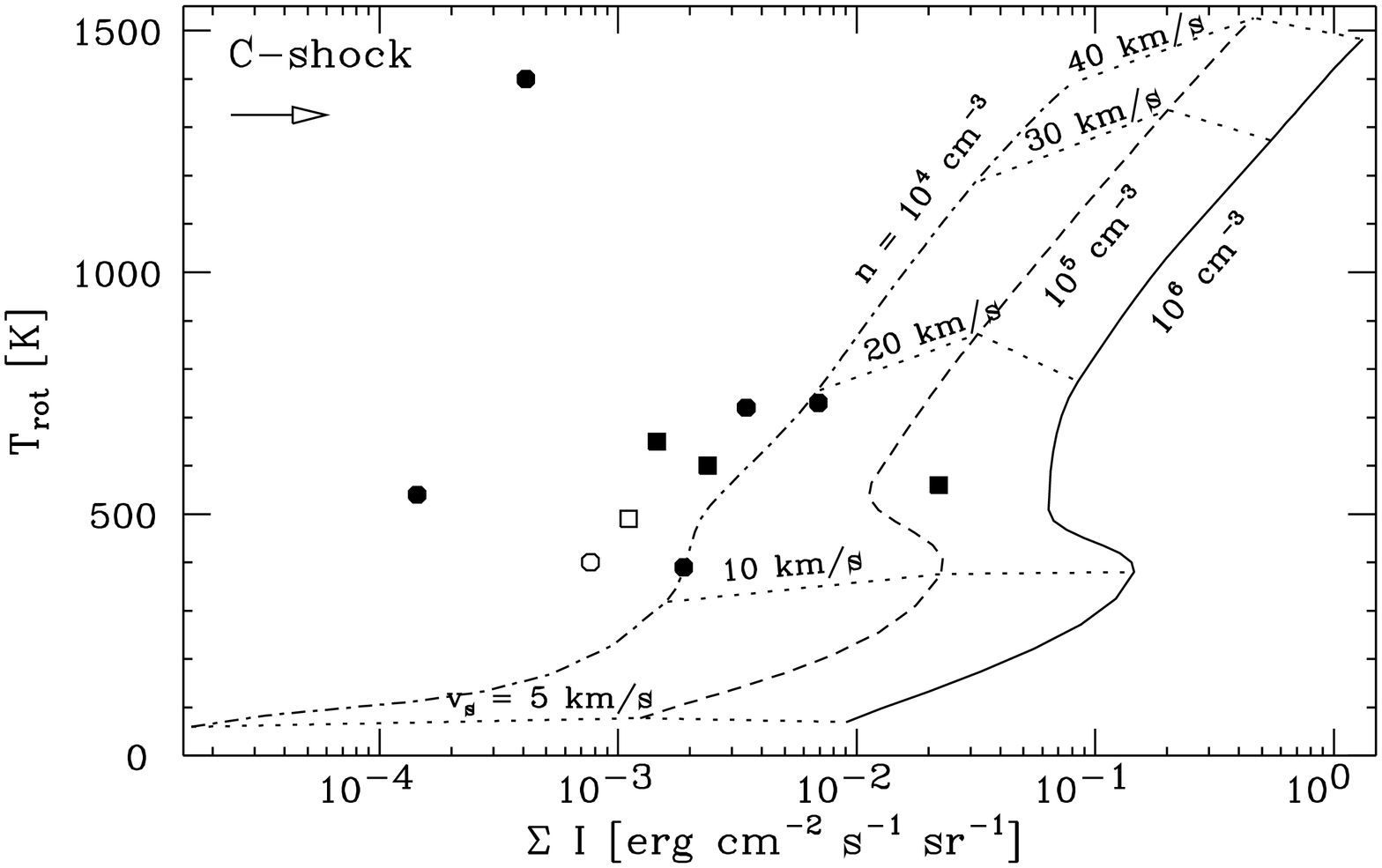,width=6.7cm,angle=90}
\hspace*{0.05cm}
\hspace*{6.7cm}}
\vspace*{-3.3cm}\hspace*{7.7cm}\parbox{6.0cm}{\small {\sc Figure 2.}
Comparison of observed H$_2$ rotational temperatures to theoretical 
relation between continuum fluxes and $T_{\rm rot}$ (PDR models) or summed 
intensity in all observed lines and $T_{\rm rot}$ (shocks). Sources which 
show PAH emission are plotted as squares. Plot symbols are filled for 
Class I sources. The arrows show the direction of beam dilution.}
\end{figure}

Until recently the study of H$_2$ in star forming regions has mainly 
concentrated on the study of the near-infrared ro-vibrational lines 
observable from the ground. However, the launch of the {\it Infrared 
Space Observatory} (ISO) has opened up the possibility to also study 
the mid-infrared pure rotational lines of H$_2$, with much lower upper 
energy levels, and directly detect the thermal emission of warm 
H$_2$ in a wide variety of sources. In these proceedings we report 
on our study of H$_2$ lines in the ISO spectra of a sample 
of 21 YSOs. We will show that emission of warm H$_2$ is common 
in the environments of intermediate- and high-mass YSOs and can  
be explained by the phenomena of shocks and PDRs outlined above.

\section{Observations and Analysis}
ISO Short Wavelength Spectrometer (SWS; 2.4 -- 45 $\mu$m) spectra were 
obtained for a sample of 21 YSOs, mostly of intermediate and high mass. 
Data were reduced in a standard fashion using calibration files 
corresponding to OLP version 7.0. In 
each object, molecular hydrogen line fluxes or upper limits (total flux 
for line with peak flux 3$\sigma$) of 0--0 S(0) to S(11), 1--0 Q(1) to 
Q(6) and 1--0 O(2) to O(7) were determined.

Pure rotational (0--0) H$_2$ emission was detected in 12 out of our 
21 sources. Ro-vibrational (1--0) H$_2$ emission was detected in 
4 sources, all of which were also detected in the pure rotational 
lines. A first inspection of our data shows that H$_2$ emission 
was only found in the vicinity of early-type ($<$ B4) stars or 
near embedded sources. Qualitatively this is in agreement with 
one would expect: the strong UV fluxes of early-type stars are expected 
to produce extended PDRs, whereas embedded YSOs are expected to drive 
strong outflows, causing a shock as the outflow hits the surrounding 
molecular cloud.

The 28.2188 $\mu$m 0--0 S(0) line was not detected. This shows directly 
that we did not detect the cool quiescent H$_2$ in the molecular cloud. A 
more qualitative analysis of our data can be made by plotting the log of
$N(\rm J)/g$, the apparent column density for a given J upper level 
divided by the statistical weight, versus the energy of the upper level. 
For the statistical weight we have assumed the high temperature 
equilibrium relative abundances of 3:1 for the ortho and para forms of
H$_2$. The resulting excitation diagrams are 
shown in Fig.~1. For comparison we also show excitation diagrams of 
three sources known to be dominated by shocks (Orion BN/KL peak 1, 
IC 443 and RCW 103) and three well-known PDRs (the Orion Bar, NGC 2023 
and S140), created using data from literature. In Fig.~1 we also show 
Boltzmann distribution fits to the low-lying pure rotational lines. 
The fact that for most sources the the points for ortho and para H$_2$ 
lie are both well fitted by this nearly straight line proves that our 
assumption on their relative abundances is correct. For a number of 
sources, the lines at higher energy levels can be seen to deviate strongly 
from the Boltzmann fit. In these cases, we have attempted to characterize 
this behaviour, which may reflect the combined effects of UV-pumped infrared 
fluorescence and the presence of a very warm, but thin, surface layer in 
a PDR and may be due to the effect of re-formation of H$_2$ in J-shocks, by 
fitting a second Boltzmann distribution to the higher energy level populations. 
The resulting excitation temperatures and derived mass of molecular 
hydrogen are also indicated in Fig.~1.

Employing predictions of H$_2$ emission from PDR, J-shock and C-shock models 
by Burton et al. (1992), Hollenbach \& McKee (1989) and Kaufman \& Neufeld 
(1996), we determined the excitation temperature $T_{\rm rot}$ from the 
low-lying pure rotational levels as a function of density $n$ and either 
incident FUV flux $G$ (in units of the average interstellar FUV field G$_0$) 
or shock velocity $v_s$ in an identical way as was done for the observations. 
The results of this procedure, plotted against $G$ (PDRs) or the total 
flux observed in all lines (shocks) are shown in Fig.~2. 
Note that there is considerable overlap between the $T_{\rm rot}$ predicted 
by PDR, J-shock and C-shock models. This means that pure rotational H$_2$ 
emission alone cannot distinguish between these mechanisms in all cases 
and additional information will be needed. However, our ISO spectra 
provide just such information. Detectable [S\,{\sc i}] 25.25~$\mu$m 
emission means that a shock must be present, whereas PAH emission 
is indicative of the presence of a PDR. The presence or absence of ionic 
lines such as [Si\,{\sc ii}] 34.82~$\mu$m can further distinguish between 
a C-shock and a J-shock. The information from these lines is used in 
Fig.~2 to distinguish between likely shocks (circles) and PDRs 
(squares).

In general, the PDR sources fall, within errors, in the parameter 
space outlined by the PDR models. Since the J-shocks only predict 
a very narrow range of $T_{\rm rot}$, only one source, AFGL 2591, 
is compatible with the observed H$_2$ emission arising in such a 
dissociative shock. All shock sources are compatible with the range 
in $T_{\rm rot}$ predicted by C-shock models. However, the detection 
of ionic lines in most of these sources shows that a J-shock component 
must be present. Most likely real astrophysical shocks are never as 
simple as the purely dissociative or non-dissociative shocks in the 
employed models, but are made up of a combination of the two, 
with the non-dissociative component dominating the H$_2$ spectrum.

\section{Conclusions} 
We have shown that pure-rotational emission from warm H$_2$ is 
readily detectable in the vicinity of intermediate- and high-mass 
YSOs and can be used to gain insight in the physical conditions 
in the circumstellar material. The main mechanisms that produce warm 
H$_2$ in these types of environments are shocks and PDRs. No 
deviations from the 3:1 ortho/para ratio of H$_2$ were found for 
either type of heating mechanism. Both shocks and PDRs show a warm 
and a hot component in H$_2$. The warm component probes the thermal 
emission from warm gas. For PDRs the hot component may reflect
the combined effects of UV-pumped infrared fluorescence and the 
presence of a thin, very warm surface layer. In shocks 
the hot H$_2$ component may be due to the re-formation of H$_2$ 
with non-zero formation energy. The warm H$_2$ component in 
shocks appears to be dominated by the non-dissociative part 
of the shock. The evolution of YSOs is expected to be from 
shock-dominated to PDR-dominated and H$_2$ may be one of 
the best tracers of the end of the outflow phase of a young 
star.

\end{document}